\documentclass[10pt,conference]{IEEEtran}
\IEEEoverridecommandlockouts
\usepackage[linesnumbered,ruled,vlined]{algorithm2e}

\usepackage{cite}
\usepackage{amsmath,amssymb,amsfonts}
\usepackage{algorithmic}
\usepackage{graphicx}
\usepackage{textcomp}
\usepackage{xcolor}
\usepackage{physics}
\usepackage{multirow}
\usepackage{enumitem}
\usepackage{mhchem}
\usepackage{booktabs}
\usepackage{subcaption}
\usepackage[linesnumbered,ruled,vlined]{algorithm2e}

\SetCommentSty{mycommfont}

\usepackage{tikz}
\usetikzlibrary{quantikz2}

\def\BibTeX{{\rm B\kern-.05em{\sc i\kern-.025em b}\kern-.08em
    T\kern-.1667em\lower.7ex\hbox{E}\kern-.125emX}}

\usepackage[bookmarks=true,breaklinks=true,colorlinks,linkcolor=blue,citecolor=blue,urlcolor=blue]{hyperref}

\usepackage{comment}
\usepackage{xspace}
\newcommand{\design}{{Qracle}\xspace}

\begin{document}

\title{Qracle: A Graph-Neural-Network-based Parameter Initializer for Variational Quantum Eigensolvers}

\author{
\begin{tabular}{ccc}
Chi Zhang & Lei Jiang & Fan Chen \\
\multicolumn{3}{c}{Indiana University Bloomington}\\
\multicolumn{3}{c}{\{czh4, jiang60, fc7\}@iu.edu} \\
\end{tabular}
}

\maketitle


\begin{abstract}
Variational Quantum Eigensolvers (VQEs) are a leading class of noisy intermediate-scale quantum (NISQ) algorithms with broad applications in quantum physics and quantum chemistry. However, as system size increases, VQE optimization is increasingly hindered by the barren plateau phenomenon, where gradients vanish and the loss function becomes trapped in local minima. While machine learning-based parameter initialization methods have been proposed to address this challenge, they often show limited effectiveness in complex VQE problems. This is primarily due to their inadequate ability to model the intricate correlations embedded in the Hamiltonian structure and the associated ansatz circuits.
In this paper, we propose \textit{Qracle}, a graph neural network (GNN)-based parameter initializer for VQEs. \textit{Qracle} systematically encodes both the Hamiltonian and the associated ansatz circuit into a unified graph representation and leverages a GNN to learn a mapping from VQE problem graphs to optimized ansatz parameters.
Compared to state-of-the-art initialization techniques, \textit{Qracle} achieves a reduction in initial loss of up to $10.86$, accelerates convergence by decreasing optimization steps by up to $64.42\%$, and improves final performance with up to a $26.43\%$ reduction in Symmetric Mean Absolute Percentage Error (SMAPE), with all results averaged over the validation set of each respective application.
The source code is publicly available at \href{https://github.com/chizhang24/Qracle}{https://github.com/chizhang24/Qracle}.
\end{abstract}


\section{Introduction}
The Variational Quantum Eigensolver (VQE)~\cite{peruzzo2014variational, mcclean2016theory} is one of the most prominent algorithms developed for noisy intermediate-scale quantum (NISQ) devices. It employs a hybrid quantum-classical optimization framework to approximate the ground-state energy of a Hamiltonian. A parameterized quantum circuit prepares trial states encoding the problem of interest, and the expectation value of the Hamiltonian is estimated via quantum measurements. A classical optimizer then updates the circuit parameters using gradients of a predefined cost function, driving convergence toward the ground state. Owing to its flexibility~\cite{kandala2017hardware}, low circuit complexity~\cite{zhang2022variational}, and resilience to quantum noise~\cite{cerezo2021variational}, VQE has emerged as a widely studied algorithm  for NISQ devices. It has been successfully applied to a range of problems, including quantum many-body physics~\cite{bharti2022noisy, bauer2020quantum, hensgens2017quantum}, molecular chemistry~\cite{li2019variational}, and other simulation tasks~\cite{tilly2022variational}, demonstrating its potential as a practical framework for hybrid quantum-classical optimization. 

Despite its promise, VQE faces a major scalability challenge: the \textit{barren plateau} phenomenon~\cite{mcclean2018barren}, where the optimization landscape flattens exponentially with problem size, leading to vanishing gradients and ineffective optimization. Among proposed mitigation strategies~\cite{larocca2024review}, parameter initialization plays a critical role by setting the initial landscape for VQE optimization and guiding it toward regions with larger gradients, thereby enhancing trainability and reducing the risk of convergence to poor local minima.
Traditional methods, such as Hartree–Fock initialization~\cite{fedorov2022vqe}, offer reasonable starting points but fail to capture electron correlation~\cite{jones2022chemistry}, limiting their effectiveness in more general correlated systems~\cite{Gulania:JCP2021}. To address these limitations, recent data-driven approaches~\cite{Miao:PRA2024, mesman2024nn, zhang2025diffusion} employ classical neural networks—including multilayer perceptrons (MLPs), autoencoders, and diffusion models—to automatically learn mappings from problem-specific features to VQE circuit parameters, leveraging the ability of machine learning to identify useful patterns from prior instances.

While machine learning-based methods offer flexibility and potential for generalization across problem instances, their performance remains limited due to their inability to construct expressive, structured embeddings that capture the full complexity of a VQE setup—including both the target Hamiltonian and the chosen ansatz. In particular, MLP-based models~\cite{Miao:PRA2024} and autoencoders~\cite{mesman2024nn} treat VQE parameters as unstructured vectors, disregarding parameter correlations and failing to exploit the underlying physical structure encoded in the Hamiltonian.
Although the diffusion-based method~\cite{zhang2025diffusion} achieves state-of-the-art (SOTA) performance, it relies on OpenCLIP~\cite{ilharco_gabriel_2021_5143773} to embed textual descriptions of quantum problems. However, since OpenCLIP is pretrained on general-purpose data and never exposed to quantum-specific content, it lacks the domain knowledge required to capture the structural nuances of VQE tasks—resulting in suboptimal embeddings and limited applicability in practical quantum settings.

In this work, we aim to address the aforementioned limitations of existing VQE parameter initialization methods. Below, we outline the core challenges and summarize our findings, proposed solutions, and contributions.

\begin{itemize}[leftmargin=*]
    \item \textbf{Prior methods are limited in performance and generalizability due to their failure to capture the structural intricacies of VQE tasks.}  
    Our preliminary study (Section~\ref{sec:prelim}) of the SOTA diffusion-based method~\cite{zhang2025diffusion} reveals significant limitations under realistic conditions. When scaling the Heisenberg XYZ dataset from 242 instances, as used in the original work, to 2,000 instances, the mean relative error (MRE) increases from 4.9\% to 8.14\%, highlighting reduced scalability. Furthermore, applying the method to the hydrogen molecule—a system characterized by a more complex Hamiltonian not considered in the original study—yields an MRE of 46.67, underscoring its limited generalization capability. We attribute these issues to the method’s inability to capture structural intricacies embedded in the Hamiltonian structure and the associated ansatz circuit. As a preliminary exploration, we na\"ively encoded both components into a unified graph and trained a baseline graph neural network (GNN) for parameter initialization. Initial results show improvements in MRE by 4.62\% and 19.93\% on the Heisenberg XYZ model and hydrogen molecule, respectively. \textit{These findings highlight the importance of structurally informed embeddings for enhancing the scalability and generality of VQE initialization.}

    \item \textbf{GNN-based methods show promise but require generalizable graph representations, sufficient data, and task-specific architectures.} 
    While graphs naturally capture relationships among complex data entities, no prior work has systematically encoded both the problem Hamiltonian and the VQE ansatz within a unified graph representation. Existing datasets are also limited—for instance, the diffusion-based method~\cite{zhang2025diffusion} focuses solely on quantum many-body Hamiltonians and includes only 242 instances—and no GNN architecture has been explicitly designed for such graph-structured VQE problems. To address these gaps, we propose \textit{Qracle} (Section~\ref{sec:design}), a general GNN-based framework for VQE parameter initialization. \textit{\underline{First}}, Qracle encodes both the Hamiltonian and ansatz into a unified graph, enabling structure-aware and context-sensitive parameter generation. \textit{\underline{Second}}, it constructs diverse datasets spanning quantum  many-body systems, quantum chemistry, and random VQE circuits, supporting both model training and standardized benchmarking. 
    \textit{\underline{Third}}, it introduces a hybrid GNN architecture that integrates multiple aggregation and update mechanisms, specifically tailored to the VQE initialization task.

    \item \textbf{Robust evaluation of VQE parameter initialization requires comprehensive metrics.} 
    We implement \design~and evaluate it against two baselines—random initialization and the SOTA diffusion-based method~\cite{zhang2025diffusion}—using the datasets constructed in this work. The evaluation covers three core aspects: initial loss, convergence efficiency, and final VQE performance, with metrics detailed in Section~\ref{sec:exp_method} and results reported in Section~\ref{sec:results}. Overall, \design consistently outperforms the SOTA diffusion-based baseline across all applications and evaluation metrics. Averaged across each application's validation set, \design achieves \textit{lower initial loss} (up to 10.86 reduction), \textit{faster convergence} (up to 64.42 fewer optimization steps), and \textit{improved final VQE performance} (up to 26.43\% reduction in Symmetric Mean Absolute Percentage Error, SMAPE), demonstrating its effectiveness as a general-purpose initializer across diverse VQE settings.   In comparison to random initialization, we observe that in a few cases, random initialization achieves comparable or even superior performance on certain metrics. This outcome reflects the stochastic nature of VQE optimization and indicates that, while \design offers robust and generalizable performance overall, there remain isolated scenarios where simpler heuristics may still perform competitively.
\end{itemize}


\vspace{0.05in}
\section{Background and Motivation}

\subsection{Key Concepts in Variational Quantum Eigensolver}
\textbf{Basics}.
The Variational Quantum Eigensolver (VQE)~\cite{peruzzo2014variational, mcclean2016theory} is a class of NISQ algorithms designed to estimate an upper bound on the ground-state energy of a target problem Hamiltonian ($\hat{H}$). 
This work focuses on general local-interaction Hamiltonians relevant to quantum many-body systems~\cite{takhtadzhan1979quantum, hensgens2017quantum, bauer2020quantum} and quantum chemistry~\cite{li2019variational, tilly2022variational}, which are commonly represented as weighted sums of Pauli strings:
\begin{equation}
\hat{H} = \sum_i \alpha_i \hat{P}_i
\label{e:vqe_ham_rep}
\vspace{-0.05in}
\end{equation}
where $\alpha_i$ are real-valued coefficients, and each $\hat{P}_i$ denotes a Kronecker product of Pauli operators $I$, $X$, $Y$, and $Z$.

\begin{figure}[t!]
\centering
\includegraphics[width=3.4in]{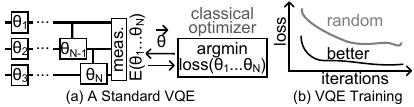}
\caption{A standard VQE and its optimization.}
\label{f:vqe_training_all}
\vspace{-0.15in}
\end{figure}

\textbf{Optimization}.
As conceptually illustrated in Figure~\ref{f:vqe_training_all}(a), a VQE typically employs a parameterized quantum circuit (i.e., ansatz)~\cite{yang2024maximizing, du2022quantum, nakaji2024generative}, which may consist of a single or multiple layers, with trainable parameters $\{\theta_i\}_{i=1}^N$ that prepare a quantum state $\ket{\psi(\theta_1, \ldots, \theta_N)}$. The expectation value of the Hamiltonian $\hat{H}$ with respect to this state is computed as:
\begin{equation}
\begin{split}
    \expval{\hat{H}} &= \expval{\hat{H}}{\psi(\theta_1, \ldots, \theta_N)} \\
    &= \sum_i \alpha_i \expval{\hat{P}_i}{\psi(\theta_1, \ldots, \theta_N)}
\end{split}
\end{equation}
To minimize $\expval{\hat{H}}$, VQE employs a classical optimizer~\cite{mcclean2016theory} in an iterative optimization process, where the loss function is defined as $f(\theta_1, \ldots, \theta_N) = \expval{\hat{H}}{\psi(\theta_1, \ldots, \theta_N)}$. 
The optimizer seeks a minimum of $f$ by updating the parameter vector
$\va{\theta} = (\theta_1, \ldots, \theta_N)$ using a gradient-based rule:
\begin{equation}
\va{\theta}^{(1)} = \va{\theta}^{(0)} - l \nabla f(\va{\theta}^{(0)})
\end{equation}
where $\va{\theta}^{(0)}$ and $\va{\theta}^{(1)}$ denote the parameters before and after the update, $l$ is the learning rate, and $\nabla f(\va{\theta}^{(0)})$ is the gradient of the objective function evaluated at $\va{\theta}^{(0)}$.

\begin{figure*}[t]\centering
\begin{subfigure}[t]{0.33\textwidth}\centering
\includegraphics[width=\linewidth]{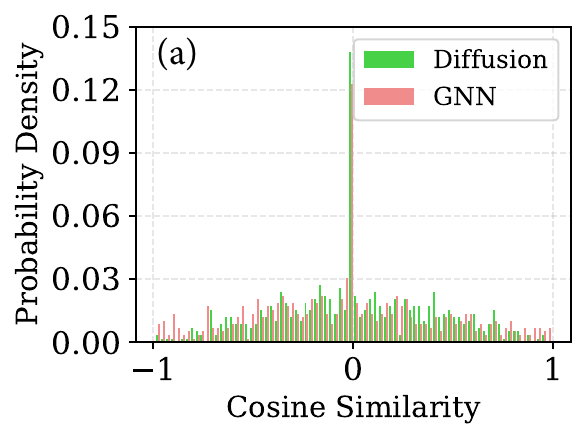}
\vspace{-0.25in}
\label{f:heisenberg_cosine_sim}
\end{subfigure}
\hspace{-0.08in}
\begin{subfigure}[t]{0.33\textwidth}\centering
\includegraphics[width=\linewidth]{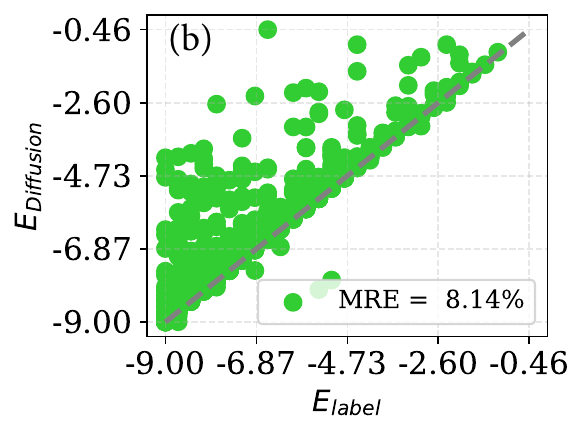} 
\vspace{-0.25in}
\label{f:heisenberg_diff_conf_mat}
\end{subfigure}
\hspace{-0.1in}
\begin{subfigure}[t]{0.33\textwidth}\centering
\includegraphics[width=\linewidth]{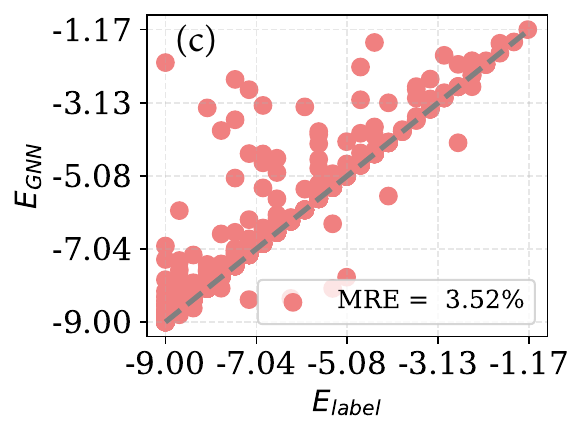} 
\vspace{-0.25in}
\label{f:heisenberg_qracle_conf_mat}
\end{subfigure}
\vspace{-0.05in}
\caption{Generative performance of diffusion- and GNN-based methods on the \textbf{\textit{Heisenberg XYZ model}}. (a) Cosine similarity distribution. (b, c) Confusion plots of predicted ground-state energies ($E_{\text{diffusion}}$, $E_{\text{GNN}}$) vs.\ ground truth ($E_{\text{label}}$).}
\label{f:heisenberg_cos_sim_conf_mat}
\vspace{-0.1in}
\end{figure*}
\begin{figure*}[t]\centering
\begin{subfigure}[t]{0.33\textwidth}\centering
\includegraphics[width=\linewidth]{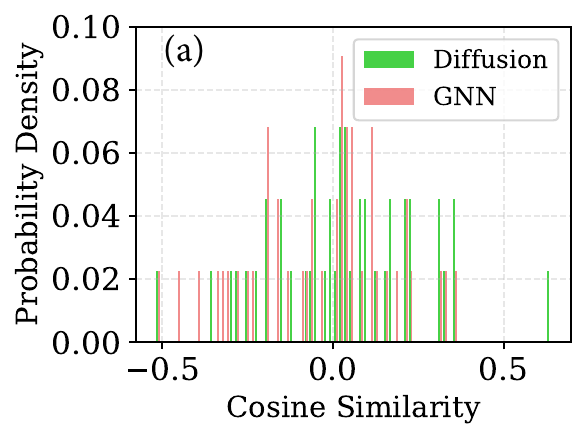}
\vspace{-0.25in}
\label{f:h2_cosine_sim}
\end{subfigure}
\hspace{-0.08in}
\begin{subfigure}[t]{0.33\textwidth}\centering
\includegraphics[width=\linewidth]{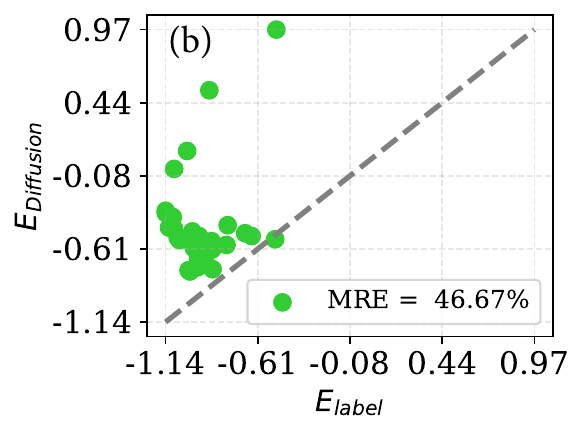} 
\vspace{-0.25in}
\label{f:h2_diff_conf_mat}
\end{subfigure}
\hspace{-0.1in}
\begin{subfigure}[t]{0.33\textwidth}\centering
\includegraphics[width=\linewidth]{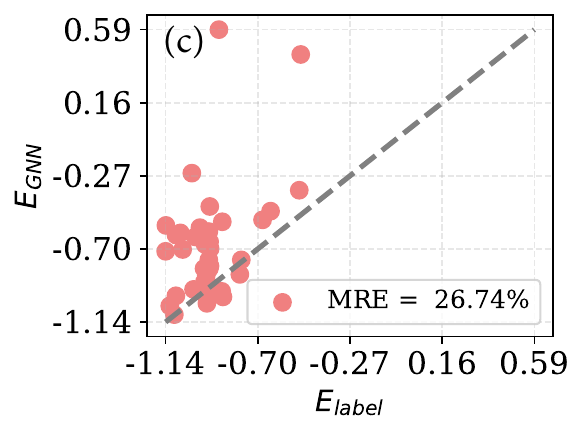} 
\vspace{-0.25in}
\label{f:h2_diff_conf_mat}
\end{subfigure}
\vspace{-0.05in}
\caption{Generative performance of diffusion- and GNN-based methods on the \textbf{\textit{Hydrogen molecule model}}. (a) Cosine similarity distribution. (b, c) Confusion plots of predicted ground-state energies ($E_{\text{diffusion}}$, $E_{\text{GNN}}$) vs.\ ground truth ($E_{\text{label}}$).}
\label{f:h2_cos_sim_conf_mat}
\vspace{-0.15in}
\end{figure*}

\textbf{Barren Plateau and Parameter Initialization}.
As the system size increases---with more qubits and greater circuit depth---the gradients of randomly initialized VQE ansatz parameters decay exponentially. This phenomenon, known as the \textit{barren plateau} problem~\cite{mcclean2018barren}, creates an excessively flat optimization landscape where gradient-based methods fail due to vanishing gradients, often trapping the optimization process in suboptimal regions and severely hindering convergence and accuracy.
To mitigate this issue, effective initialization strategies for VQE ansatzes have become essential~\cite{larocca2024review}. As shown in Figure~\ref{f:vqe_training_all}(b), carefully chosen initial parameters can preserve non-negligible gradients, enabling more stable and informative updates during early optimization. These strategies not only lower the initial cost but also accelerate convergence and enhance overall VQE performance.

\subsection{Graph Neural Networks}
A graph is defined as $\mathcal{G} = (\mathcal{V}, \mathcal{E}, \mathcal{W}, \mathcal{X})$, where $\mathcal{V}$ is the set of $n$ nodes and $\mathcal{E}$ the set of $m$ edges, while $\mathcal{W}$ is the set of edge weights and $\mathcal{X}$ the set of node features. A directed edge from node $v_i$ to node $v_j$ is denoted as $e_{ij} = (v_i, v_j) \in \mathcal{E}$. The graph structure encoded by $(\mathcal{V}, \mathcal{E}, \mathcal{W})$ can be equivalently represented by an adjacency matrix $W \in \mathbb{R}^{n \times n}$, where $W_{ij}$ encodes the presence or weight of an edge. Each node $v_i$ is typically associated with a feature vector $\mathbf{x}_i \in \mathbb{R}^d$, and the full set of node features forms a matrix $\mathcal{X} \in \mathbb{R}^{n \times d}$.

Graph Neural Networks (GNNs)~\cite{kipf2016semi,Velickovic:ICLR2018,xu2018how,hamilton2017inductive} learn from graph-structured data by iteratively propagating information across edges and aggregating neighborhood features to update node representations. These learned embeddings serve as the foundation for a variety of downstream tasks. A widely adopted formulation is the graph convolutional paradigm, which generalizes the notion of convolution to irregular graph domains, facilitating localized and hierarchical feature learning.
Given input features from the previous layer $\mathbf{h}_u^{(k-1)}$, a graph convolutional layer (GCN)~\cite{kipf2016semi} updates the representation of each node by aggregating information from its local neighborhood as follows:
\begin{equation}
\resizebox{.9\hsize}{!}{
$\mathbf{h}_v^{(k)} = \sigma \left( \left( \sum_{u \in N(v) \cup \{v\}} \frac{\mathbf{h}_u^{(k-1)}}{\sqrt{d(v) d(u)}} \right) \cdot W^{(k)} + b^{(k)} \right)$
}
\end{equation}
where $N(v)$ denotes the set of neighbors of $v$, and $d(v)$ represents the degree of $v$. The weight matrix $W^{(k)}$ and bias vector $b^{(k)}$ are trainable parameters of the $k$-th layer, and $\sigma^{(k)}$ is a non-linear activation function (e.g., ReLU). 
The inclusion of node $v$ in its own neighborhood ensures that the model incorporates self-loops, which is equivalent to modifying the adjacency matrix as $\tilde{A} = A + I$. This ensures that each node retains and transforms its own features while incorporating context from adjacent nodes, enabling the model to learn expressive and structure-aware representations.

\section{Related Work and Preliminary Study}
\label{sec:prelim}

\subsection{Related Work}
The Hartree-Fock method~\cite{fedorov2022vqe} is commonly used to initialize VQE parameters. However, as a mean-field approximation, it neglects higher-order electron correlations~\cite{jones2022chemistry}, limiting its effectiveness for strongly correlated Fermion systems~\cite{Gulania:JCP2021}. 
To address this, recent studies have explored classical neural networks~\cite{mesman2024nn,Miao:PRA2024,zhang2025diffusion} that learn parameter distributions from optimized ansatzes to improve generalization. Nevertheless, these methods fail to incorporate the underlying Hamiltonian structure, which is essential in  VQE algorithms. The MLP-based approach in~\cite{Miao:PRA2024} embeds Hamiltonian coefficients and VQE parameters without modeling parameter correlations, restricting applicability to simple problems such as 1D Ising models. The quantum autoencoder framework in~\cite{mesman2024nn} reduces the parameter space in more complex settings but still relies on MLPs that overlook key dependencies. A recent diffusion-based method~\cite{zhang2025diffusion} leverages OpenCLIP~\cite{ilharco_gabriel_2021_5143773} to encode Hamiltonian descriptions—e.g., \texttt{"J and h: [1,1,3...]
Pauli item: [XXI,IZZ...]"}—as conditioning inputs for the diffusion model training. However, since OpenCLIP is trained on general-purpose data and lacks quantum domain knowledge, it cannot adequately capture Hamiltonian structure, limiting its effectiveness for VQE optimization.

\subsection{Preliminary Study}
We implemented the diffusion-based method from~\cite{zhang2025diffusion} and extended its evaluation using a larger dataset of 2,000 Heisenberg XYZ instances, compared to the original 242. 
We encode Hamiltonians as VQE parameter vectors (see Section~\ref{sec:design}) and represent additional information—such as qubit count and Hamiltonian configuration—as node features. We implement a na\"ive GNN model consisting of two GCN layers followed by an MLP to learn the mapping from graphs to VQE parameter vectors. Details of the experimental setup are provided in Section~\ref{sec:exp_method}.
We adopt the same metrics as~\cite{zhang2025diffusion}: cosine similarity between generated and reference parameters, and mean relative error (MRE) of the VQE-computed ground-state energy. 
As shown in Fig.~\ref{f:heisenberg_cos_sim_conf_mat}(a), both models produced parameter vectors distinct from the labels. However, the diffusion model’s MRE increased from 4.9\% to 8.14\% on the larger dataset, indicating reduced scalability. In contrast, the GNN model achieved a lower MRE of 3.52\%.
To assess generalization, we tested both models on the hydrogen molecule, which involves a more complex Hamiltonian not considered in the original study. As shown in Fig.~\ref{f:h2_cos_sim_conf_mat}, the diffusion model yielded a high MRE of 46.67\%, highlighting poor adaptability to out-of-distribution inputs. The GNN model, however, significantly reduced the MRE to 26.74\%, suggesting improved capacity to capture structural dependencies between the Hamiltonian and the VQE ansatz. These results motivate further exploration of GNN-based parameter initialization.

\section{\design}
\label{sec:design}
In this section, we introduce \textit{Qracle}, a GNN-based initializer for VQE parameters.
Qracle first transforms Pauli-operator Hamiltonians into adjacency matrices, enabling a natural and expressive graph representation.
We then describe three representative applications, detailing how application-specific information is encoded, the structure of the circuit ansatz, and the dataset construction process. 
Finally, we present the GNN architecture used in Qracle, which integrates multiple aggregation and update mechanisms to generate high-quality initialization parameters for previously unseen VQE ansatzes.

\begin{figure}[t!]
\centering
\includegraphics[width=3in]{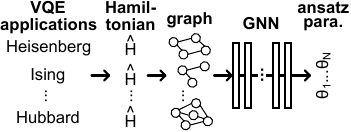}
\caption{An overview of Qracle.}
\label{f:vqe_over_all}
\vspace{8pt}
\end{figure}

\subsection{Graph Representation of Quantum Hamiltonians}
\label{subsec:graph_representation}

\setlength{\textfloatsep}{0pt}
\begin{algorithm}[b]
\DontPrintSemicolon
\caption{Convert Hamiltonian $\hat{H}$ to Graph $\mathcal{G}$}\label{alg:hamiltonian_to_graph}
\SetKwInOut{Input}{Input}\SetKwInOut{Output}{Output}

\Input{Hamiltonian $\hat{H}$ of an $n$-qubit system}
\Output{Directed weighted graph $\mathcal{G} = (\mathcal{V}, \mathcal{E}, \mathcal{W}, \mathcal{X})$}

\If{$\hat{H}$ not in Pauli-operator form in Eq.~\ref{e:explicit_pauli_string}}{
    Apply transformations (e.g., Jordan--Wigner, Bravyi--Kitaev) to convert $\hat{H}$
}

Compute matrix $H_{pq}$ of $\hat{H}$ using Eq.~\ref{e:h_qp_value} \;

$\mathcal{V}, \mathcal{E}, \mathcal{W}, \mathcal{X} \gets \emptyset$ \tcp*{Initialize graph components}

\For{$p \gets 1$ \KwTo $2^n$}{
    $v_p \gets$ vertex for $p$; $\mathcal{V}.\texttt{add}(v_p)$ \;
    $\mathbf{x}_p \gets$ node feature for $v_p$; $\mathcal{X}.\texttt{add}(\mathbf{x}_p)$ \;
    
    \For{$q \gets 1$ \KwTo $2^n$}{
        \If{$H_{pq} \neq 0$}{
            $\mathcal{E}.\texttt{add}((v_p \rightarrow v_q))$ ;  $\mathcal{W}.\texttt{add}(H_{pq})$  \;
        }
    }
}
$\mathcal{G} \gets (\mathcal{V}, \mathcal{E}, \mathcal{W}, \mathcal{X})$ \;
\KwRet{$\mathcal{G}$}
\end{algorithm}

For an $n$-qubit system, the Hamiltonian in Eq.~\eqref{e:vqe_ham_rep} can be expanded as a linear combination of Pauli strings:
\begin{equation}
\hat{H} = \sum_{\boldsymbol{P} \in \mathcal{P}^{\otimes n}} \alpha_{\boldsymbol{P}}\, P_0 \otimes P_1 \otimes \cdots \otimes P_{n-1}
\label{e:explicit_pauli_string}
\end{equation}
where each $P_k \in \{I, X, Y, Z\}$ is a Pauli operator acting on the $k$-th qubit, and $\alpha_{\boldsymbol{P}} \in \mathbb{C}$ is its coefficient.  The set $\mathcal{P}^{\otimes n}$ comprises all $4^n$ possible Kronecker products of $n$ Pauli operators. Since each $P_k$ is $2 \times 2$, the resulting Hamiltonian $\hat{H}$ is a $2^n \times 2^n$ Hermitian matrix, given by:
\begin{equation}
\hat{H} = \begin{pmatrix}
H_{11} & H_{12} & \cdots & H_{1, 2^n} \\
H_{21} & H_{22} & \cdots & H_{2, 2^n} \\
\vdots & \vdots & \ddots & \vdots \\
H_{2^n, 1} & H_{2^n, 2} & \cdots & H_{2^n, 2^n}
\end{pmatrix}
\label{e:ham_all_matrix}
\end{equation}
where $H_{pq}$ denotes the $(p, q)$-th entry of the matrix,  with $p, q \in \{1, \ldots, 2^n\}$. By substituting $P_k \in \{ I, X, Y, Z\}$  into Eq.~\eqref{e:explicit_pauli_string} and explicitly expanding the Kronecker product, each matrix element $H_{pq}$ can be computed as:
\vspace{-4pt}
\begin{equation}
\begin{aligned}
H_{pq} &= \sum_{\boldsymbol{P} \in \mathcal{P}^{\otimes n}} \alpha_{\boldsymbol{P}} \prod_{k=0}^{n-2} (P_k)_{\lceil p / 2^{n-k-1} \rceil,\ \lceil q / 2^{n-k-1} \rceil} \\
&\quad \times (P_{n-1})_{(p-1)\bmod 2 + 1,\ (q-1)\bmod 2 + 1}
\end{aligned}
\label{e:h_qp_value}
\end{equation}
where $\lceil \cdot \rceil$ denotes the ceiling function, and $\bmod$ represents the modulo operation. This explicit mathematical transformation allows any Pauli-operator-based Hamiltonian to be represented as the adjacency matrix of a weighted, directed graph with  $\leq 2^n$ nodes. Such a graph-based representation facilitates the application of graph-theoretic and geometric techniques in quantum algorithms and machine learning.

\begin{table*}[t!]
\footnotesize
\caption{
Summary of the constructed quantum graph-to-vector datasets. 
Each graph represents a problem-specific Hamiltonian, and the corresponding label vector consists of optimized parameters obtained from a Variational Quantum Eigensolver (VQE) trained to approximate the ground state of that Hamiltonian.
}
\label{tab:quantum_dataset_summary}
\centering
\setlength{\tabcolsep}{1.8pt}
\begin{tabular}{|c|c|c|c||c|c|c|c|}
\hline

\multirow{2}{*}{\textbf{Application}}   
& \multicolumn{3}{c||}{\textbf{Dataset Configuration}} 
& \multicolumn{4}{c|}{\textbf{Ansatz Configuration}} 
\\\cline{2-8}
& \textbf{\# Graphs} 
& \textbf{Label Dim.} 
& \textbf{Node Features} 
& \textbf{\# Qubits ($n_q$)} 
& \textbf{\# Layers ($l$)} 
& \textbf{Params/Qubit ($n_p$)}  
& \textbf{Backend} 
\\ \hline\hline

\textbf{Heisenberg XYZ} 
& 2,000 
& $1 \times 8$ 
& (index, $n_q$, $J_1$, $J_2$, $J_3$) 
& 4 
& 1 
& 2 
& \texttt{PennyLane } \\
\textbf{2D Transverse-Field Ising} 
& 1,000 
& $1 \times 16$ 
& (index, $n_q$, $J$, $\mu$) 
& 8 
& 1 
& 2
& \texttt{PennyLane } \\
\textbf{Fermi--Hubbard} 
& 1,000 
& $1 \times 16$ 
& (index, $n_q$, $t$, $U$) 
& 8 
& 1 
& 2 
& \texttt{PennyLane } \\
\textbf{\ce{H2} Molecules} 
& 150 
& $1 \times 24$ 
& (index, $n_q$, atom coordinates) 
& 4 
& 2 
& 3
& \texttt{PennyLane } \\
\textbf{Random VQE} 
& 2,800 
& $1 \times 48$ 
& (index, $n_q$) 
& 4 
& 2 
& 6 
& \texttt{TorchQuantum} \\
\hline
\end{tabular}
\end{table*}
\begin{figure*}[t!]
\centering
\includegraphics[width=.98\linewidth]{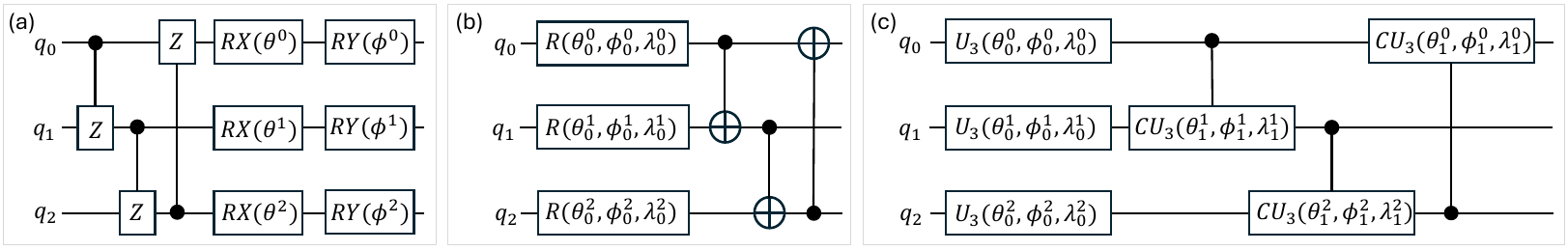}
\caption{Ansatz used for (a) many-body systems~\cite{lee2025q, zhangEscapingBarrenPlateau2022}, (b) \ce{H2} molecule hamiltonian~\cite{schuldCircuitcentricQuantumClassifiers2020, zhangEscapingBarrenPlateau2022}, and (c) Random VQE~\cite{hanruiwang2022quantumnas}.}
\label{f:Ansatz}
\vspace{-0.1in}
\end{figure*}

To illustrate this transformation, Algorithm~\ref{alg:hamiltonian_to_graph} constructs the graph by assigning a vertex to each quantum basis state (line 6), attaching the corresponding node feature (line 7), and adding directed edges weighted by the nonzero elements of the Hamiltonian matrix (line 10).


\subsection{Representative Applications and Dataset Construction}
To evaluate the effectiveness and flexibility of the~\design~framework, we consider the quantum many-body physics Hamiltonians used in prior work~\cite{zhang2025diffusion}, and further incorporate a quantum chemistry Hamiltonian~\cite{li2019variational} as well as generic random VQE instances~\cite{liQASMBenchLowLevelQuantum2023}. For each problem domain, we describe the corresponding Hamiltonians—including their conversion to Pauli string form when not originally expressed as such—and explain how application-specific information is encoded as graph node attributes. We also provide details of the dataset construction process for each case, with an overall summary presented in Table~\ref{tab:quantum_dataset_summary}.

\subsubsection{Quantum Many--Body Systems}
Similar to the SOTA diffusion-based model~\cite{zhang2025diffusion}, we study a class of quantum many-body systems, including the Heisenberg XYZ model~\cite{takhtadzhan1979quantum}, the two-dimensional (2D) transverse-field Ising model~\cite{bauer2020quantum}, and the spinless Fermi–Hubbard model~\cite{hensgens2017quantum}. 
Their Hamiltonians can be expressed in a unified form:
\begin{equation}\label{e:unified_many_body_hamiltonian}
    \hat{H} = \sum_{(i, j)} J_{ij} \hat{O}_i \hat{O}_j + \sum_{i} h_i \hat{K}_i
\end{equation}
where $(i, j)$ denotes nearest-neighbor interactions on a lattice, $J_{ij}$ represents the interaction strength between sites $i$ and $j$, $h_i$ is the local field at site $i$, and $\hat{O}_i$ and $\hat{K}_i$ denote the interaction and potential operators, respectively, whose explicit forms depend on the specific model. 
In the following, we provide the explicit forms of the Hamiltonians for each model along with details of their corresponding dataset construction.

\begin{itemize}[leftmargin=*]
    \item \textbf{Heisenberg XYZ Dataset}. 
    The specific Hamiltonian for the Heisenberg XYZ model is given in Eq.~\eqref{e:Heisenberg_XYZ}, where $X_i$, $Y_i$, and $Z_i$ denote the Pauli $X$, $Y$, $Z$ operators acting on the $i$-th qubit, respectively.
    \vspace{-4pt}
    \begin{equation}\label{e:Heisenberg_XYZ}
    \hat{H} = \sum_{i=0}^{n-1} (J_1 X_i X_{i+1} + J_2 Y_i Y_{i+1} + J_3 Z_i Z_{i+1})
    \end{equation}
    To construct the dataset for the Heisenberg XYZ model, we generate 2000 tuples of coupling constants \( (J_1, J_2, J_3) \), where each \( J_i \in [-3, 3] \) is sampled on a uniform grid with spacing \(0.1\). We select 1000 distinct configurations to define individual Hamiltonians based on Eq.~\eqref{e:Heisenberg_XYZ}. Each Hamiltonian is constructed using \texttt{PennyLane} and represented as a graph.
    For each Hamiltonian, VQE optimization is performed using the ansatz circuit~\cite{lee2025q, zhangEscapingBarrenPlateau2022} in Fig.\ref{f:Ansatz}(a), with \( n_q = 4 \) qubits and a single layer (\( l = 1 \)). The ansatz contains \( l \times 2 \times n_q = 8 \) trainable parameters.
    Graph node attributes are encoded as a \( 1 \times 5 \) vector comprising the tuple (index, \(n_q\), \(J_1\), \(J_2\), \(J_3\)). Each VQE is optimized using the \texttt{Adam} optimizer with a learning rate of \(10^{-3}\) over 2000 steps. For each run, we record the complete loss history and the final optimized parameters for downstream analysis.

    \item \textbf{2D Transverse--field Ising Dataset}. 
    The Hamiltonian for the 2D Ising model is given in Eq.~\eqref{e:Ising_model_simplified}, where $j$ is the interaction strength between neighboring spins, and $\mu$ is the transverse field strength.
    \begin{equation}\label{e:Ising_model_simplified}
    \hat{H} = -j \sum_{(i, j)} Z_i Z_j - \mu \sum_i Z_i
    \end{equation}
    To construct the dataset for the 2D Ising model, we sample 1000 pairs of coupling constants \( (j, \mu) \) for the Hamiltonian defined in Eq.~\eqref{e:Ising_model_simplified}, where \( j, \mu \in [0, 5] \) are drawn from a uniform grid with spacing \(0.1\). The corresponding Hamiltonians are constructed using \texttt{PennyLane} and represented as graphs. For each Hamiltonian, we perform VQE optimization using the ansatz circuit shown in Fig.~\ref{f:Ansatz}(a), with \( n_q = 8 \) qubits and a single layer (\( l = 1 \)). The ansatz comprises \( l \times 2 \times n_q = 16 \) trainable parameters. Graph node attributes are encoded as a \( 1 \times 4 \) vector representing the tuple (index, \(n_q\), \(j\), \(\mu\)). The VQE optimization follows the same configuration used for the Heisenberg XYZ model.

    \item \textbf{Fermi--Hubbard Dataset}. 
    The Hamiltonian for the Fermi--Hubbard model is given in Eq.~\eqref{e:fh_hamiltonian}, where $\hat{c}^\dagger_i$ and $\hat{c}_i$ are fermionic creation and annihilation operators at site $i$, $\hat{n}_i = \hat{c}^\dagger_i \hat{c}_i$ is the number operator, $t$ denotes the hopping amplitude, and $U$ the nearest-neighbor interaction strength.
    To apply VQE, the Hamiltonian must be mapped to a qubit representation. The Jordan–Wigner transformation accomplishes this by expressing fermionic operators in terms of Pauli matrices. For \( n \) spin orbitals, the transformation is defined in Equations~\ref{e:Jordan-Wigner-1}–\ref{e:Jordan-Wigner-2}, where \( X_j \), \( Y_j \), and \( Z_j \) are Pauli operators on qubit \( j \). Substituting into Eq.~\eqref{e:fh_hamiltonian} yields the qubit Hamiltonian in Pauli operator form shown in Eq.~\eqref{e:explicit_pauli_string}.
    \vspace{-0.1in}
    \begin{equation}\label{e:fh_hamiltonian}
    \hat{H} = -t \sum_{i=0}^{n-1} (\hat{c}^\dagger_i \hat{c}_{i+1} + \hat{c}^\dagger_{i+1} \hat{c}_i) + U \sum_{i=0}^{n-1} \hat{n}_i \hat{n}_{i+1}
    \vspace{-0.1in}
    \end{equation}
    \vspace{-0.1in}
    \begin{align}
    \hat{c}_j &= I^{\otimes j} \otimes \left( \frac{X_j + i Y_j}{2} \right) \otimes Z^{\otimes (n - j - 1)} \label{e:Jordan-Wigner-1} \\
    \hat{c}_j^\dagger &= I^{\otimes j} \otimes \left( \frac{X_j - i Y_j}{2} \right) \otimes Z^{\otimes (n - j - 1)} \label{e:Jordan-Wigner-2}
    \vspace{-0.1in}
    \end{align}    
   To construct the dataset for the Fermi--Hubbard model, we generate 1000 tuples of coupling constants \( (t, U) \) for the Hamiltonian in Equation~\eqref{e:fh_hamiltonian}, where \( t, U \in [0, 5] \) are sampled on a uniform grid with spacing \(0.1\). The corresponding Hamiltonians are constructed using \texttt{PennyLane} and represented as graphs. 
    For each Hamiltonian, we perform VQE optimization using the ansatz circuit shown in Fig.~\ref{f:Ansatz}(a), with \( n_q = 8 \) qubits and a single layer (\( l = 1 \)). The ansatz comprises \( l \times 2 \times n_q = 16 \) trainable parameters. Graph node attributes are encoded as a \( 1 \times 4 \) vector representing the tuple (index, $n_q$, $t$, $U$). The VQE optimization follows the same configuration used for the Heisenberg XYZ model.
\end{itemize}

\subsubsection{Quantum Chemistry}
We consider quantum chemistry as a target application, where the objective is to compute molecular ground-state energies. The key quantity is the electronic structure Hamiltonian, typically expressed in second quantization, as shown in Eq.~\eqref{e:molecule_hamiltonian}, where \( \hat{c}^\dagger_i \) and \( \hat{c}_j \) are fermionic creation and annihilation operators, and \( h_{ij} \) and \( h_{ijkl} \) are one- and two-electron integrals derived from a chosen basis set. Applying the Jordan–Wigner transformation to Eq.~\eqref{e:molecule_hamiltonian} produces a qubit Hamiltonian in terms of Pauli matrices, consistent with the general form in Eq.~\eqref{e:explicit_pauli_string}, and directly compatible with our framework.
\begin{equation}\label{e:molecule_hamiltonian}
    \hat{H} = \sum_{i, j} h_{ij} \hat{c}^\dagger_{i} \hat{c}_j + \frac{1}{2} \sum_{i, j, k, l} h_{ijkl} \hat{c}^\dagger_{i} \hat{c}^\dagger_{j} \hat{c}_k \hat{c}_l
\end{equation}
  
\textbf{\ce{H2} Molecule Dataset}.
The Hamiltonian of a molecular system depends on its geometric configuration (e.g., bond length, bond angle), and its complexity increases rapidly with molecular size. To keep the problem tractable, we focus on the \ce{H2} molecule, which consists of only two protons, and construct our dataset accordingly. Under the Born–Oppenheimer approximation~\cite{mcquarriePhysicalChemistryMolecular200}, the Hamiltonian of \ce{H2} depends solely on the bond length.
We generate a set of \ce{H2} Hamiltonians using \texttt{PennyLane} by varying the bond length from 0.5~\AA{} to 4.97~\AA{} in increments of 0.03~\AA{}, yielding 150 unique configurations. This range covers a substantial portion of the \ce{H2} entries in the \texttt{PennyLane} quantum chemistry dataset~\cite{Utkarsh2023Chemistry}. Each Hamiltonian is represented in the STO–3G basis and transformed into spin-operator form via the Jordan–Wigner mapping, resulting in Pauli string expressions as shown in Eq.~\eqref{e:explicit_pauli_string}, and is further constructed using \texttt{PennyLane} and represented as a graph.     
For each Hamiltonian, we perform VQE using a standard hardware-efficient ansatz~\cite{schuldCircuitcentricQuantumClassifiers2020, zhangEscapingBarrenPlateau2022} widely adopted in molecular simulations, illustrated in Fig.~\ref{f:Ansatz}(b), with \( n_q = 4 \) qubits and two layers (\( l = 2 \)). The ansatz comprises \( l \times 3 \times n_q = 24 \) trainable parameters. Graph node attributes are encoded as a \( 1 \times 3 \) vector representing the tuple (index, $n_q$, atom coordinates) . The VQE optimization  is performed using the \texttt{Adam} optimizer with a learning rate of $10^{-3}$ over 2000 steps. We record the full loss trajectory and final optimized parameters for each VQE run.

\subsubsection{Random VQE}
Randomly generated Hamiltonians are commonly used as benchmarks~\cite{liQASMBenchLowLevelQuantum2023}  to evaluate VQE algorithms. These Hamiltonians, defined by Eq.~\eqref{e:explicit_pauli_string}, are created by randomly choosing coefficients and Pauli operators. The inherent randomness ensures an unbiased and diverse testing environment, mitigating structural biases typical of Hamiltonians from specific physical or molecular systems. Thus, these benchmarks enable comprehensive assessments of VQE performance, effectively probing key algorithmic traits.

\textbf{Random VQE Dataset}.
To construct the dataset for the Random VQE model, we generate 2,800 random Hamiltonians using \texttt{QASMBench}~\cite{liQASMBenchLowLevelQuantum2023}. Each Hamiltonian is built with \texttt{TorchQuantum}~\cite{hanruiwang2022quantumnas} and represented as a graph.
For each Hamiltonian, we implement the VQE circuit using the standard ansatz from~\cite{hanruiwang2022quantumnas}, as shown in Figure~\ref{f:Ansatz}(c), with \( n_q = 4 \) qubits and two layers (\( l = 2 \)). The ansatz comprises \( l \times 6 \times n_q = 48 \) trainable parameters. Graph node attributes are encoded as a \( 1 \times 2 \) vector containing the tuple (index, \(n_q\)).
VQE optimization is performed using the \texttt{Adam} optimizer with a learning rate of \(5 \times 10^{-3}\), weight decay of \(1 \times 10^{-4}\), and the \texttt{CosineAnnealingLR} scheduler. For each instance in the dataset, we record the full loss history and final optimized parameters for downstream analysis.

\subsection{GNN Architecture}
\label{subsec:gnn_model}
By encoding both the Hamiltonian structure and ansatz configuration into graph representations, we design a GNN architecture that learns expressive embeddings to map target Hamiltonians and VQE ansatz circuits to optimized parameter values. Leveraging the localized correlations commonly present in quantum systems, we incorporate graph convolutional layers to iteratively aggregate neighborhood information. To capture long-range dependencies and adaptively modulate the influence of neighboring nodes, attention mechanisms are further integrated into the model. We next present the message-passing formulation and layer design of our GNN, followed by an overview of the architecture and implementation details.

\textbf{Message-Passing}.
All GNN architectures are fundamentally based on a message-passing mechanism~\cite{kipf2016semi}, in which each node aggregates information from its neighbors and updates its own representation accordingly. This process can be mathematically expressed as:
\begin{equation}
\resizebox{.88\hsize}{!}{
$V_i^{l} = \text{UPDATE} \left( V_i^{l-1}, \text{AGGREGATE} \left( V_j^{l-1} : j \in N(i) \right) \right)$}
\label{eq:graph_message_passing}
\end{equation}
where $V_i^{l}$ represents the feature vector of node $i$ at the $l$-th iteration, and $N(i)$ denotes the set of neighboring nodes adjacent to node $i$. The \text{AGGREGATE} function collects and combines features from neighboring nodes, while the \text{UPDATE} function refines the node representation. Through iterative aggregation, the GNN effectively captures the topological structure of the graph within a node’s representation.

\textbf{GNN Layers}.
\textsc{Qracle} leverages Graph Convolutional Networks (GCNs)\cite{kipf2016semi} and Graph Attention Networks (GATs)\cite{Velickovic:ICLR2018} to process the graph-structured representations of Hamiltonians. Each GNN layer architecture adopts a distinct strategy for information aggregation and representation update:
\begin{itemize}[leftmargin=*]
\item \textit{GCN}: GCNs~\cite{kipf2016semi} propagate information across nodes by updating node representations through a neighborhood aggregation scheme, as shown in Eq.~\eqref{eq:graph_gcn_update}. In this formulation, $\deg_i$ denotes the degree of node $i$ in the adjacency matrix, and $W$ is the learnable weight matrix.
\begin{equation}
V_i^{l} = \text{ReLU} \left( W^{l-1} \frac{1}{\deg_i} \sum_{j \in N(i)} V_j^{l-1} \right)
\label{eq:graph_gcn_update}
\end{equation}

\item \textit{GAT}: GATs~\cite{Velickovic:ICLR2018} introduce attention mechanisms to assign varying levels of importance to neighboring nodes, enabling adaptive and context-aware feature aggregation. The final layer update in GAT is shown in Eq.~\eqref{eq:graph_gat_update}, where $K$ is the number of attention heads, $W^k$ is the learnable weight matrix for the $k$-th head, and $e_{i,j}^{k}$ represents the attention coefficient between nodes $i$ and $j$ for the $k$-th head.
\begin{equation}
V_i^{l} = \alpha \left( \frac{1}{K} \sum_{k=1}^{K} \sum_{j \in N(i)} e_{i,j}^{k} W^{k} V_j \right)
\label{eq:graph_gat_update}
\end{equation}
\end{itemize}

\textbf{Overall Architecture}. 
Our GNN model is implemented using the \texttt{PyTorch Geometric} library and comprises two \texttt{GCNConv} layers, three \texttt{GATConv} layers, and two multilayer perceptron (MLP) layers. Each layer is followed by a \texttt{ReLU} activation function. The \texttt{GCNConv} layers use a hidden dimension of 256, the \texttt{GATConv} layers 512, and the MLP layers 1024.
The model takes a graph representation as input, and the final MLP layer maps the learned embedding to the VQE parameter vector space.
Training is performed using mean squared error loss (\texttt{MSELoss}) and optimized with the Adam optimizer with a learning rate of $1\times10^{-3}$. All training and evaluation experiments are conducted on an NVIDIA RTX 3090 GPU.

\section{Experimental Methodology}
\label{sec:exp_method}

\textbf{Datasets}.
We conduct the evaluation using five datasets spanning various application domains, including many-body physics, quantum chemistry, and randomly generated VQEs, as summarized in Table~\ref{tab:quantum_dataset_summary}. Each dataset is split into 70\% training and 30\% testing, and this partitioning is consistently applied across all evaluation methods.
Note that the graph dataset configurations in Table~\ref{tab:quantum_dataset_summary} are used to train the~\design~model. For compatibility with the diffusion-based initializer from~\cite{zhang2025diffusion}, we follow their original setup: each Hamiltonian description (e.g., \texttt{"VQE Hamiltonian=0.5*IZIZ+2.0*ZZZI"}) is encoded into a vector representation using OpenCLIP, resulting in a \texttt{text\_prompt} embedding. A latent diffusion model is then trained on paired data consisting of (\texttt{text\_prompt}, \texttt{opt\_param}).

\begin{table}[t!]
\footnotesize
\caption{Overall comparison of VQE initialization schemes. Here, $\Delta$ denotes the absolute improvement of \design over the diffusion-based baseline~\cite{zhang2025diffusion}, while \textcolor{blue}{blue} highlights indicate cases where random initialization achieves comparable or superior performance to \design~on specific metrics. All values are averaged over each application's validation set.}
\label{tab:eval_metrics}
\centering
\setlength{\tabcolsep}{3pt}
\begin{tabular}{|c|c|c|c|c|}
\hline
\textbf{Applications} 
& \textbf{Schemes} 
& \textbf{Initial Loss} 
& \textbf{SMAPE} 
& \textbf{Convergence Steps} \\\hline\hline
\multirow{4}{*}{\textbf{Heisenberg XYZ}}
& Random    &0.06   & 8.96\% & 266.93  \\
& Diffusion  &-0.16  &9.77\%  &251.89  \\
& Qracle    &-3.11  & 4.09\%  & 187.47 \\
& \textbf{$\Delta$}  
&\color{red}{2.95$\downarrow$}  
&\color{red}{5.68\%$\downarrow$}   
&\color{red}{64.42$\downarrow$} \\\hline
\multirow{4}{*}{\textbf{2D Ising}}
& Random    &-3.23  &2.84\%  & 235.93 \\
& Diffusion  &-12.18  &2.27\%   &180.83 \\
& Qracle    &-23.04  &0.39\%   &158.50 \\
& \textbf{$\Delta$}  
&\color{red}{10.86$\downarrow$}  
&\color{red}{1.88\%$\downarrow$}  
&\color{red}{22.33$\downarrow$} \\\hline
\multirow{4}{*}{\textbf{Fermi-Hubbard}}
& Random     & 1.01  &\textcolor{blue}{13.54\%} &218.46 \\
& Diffusion  &0.47  & 21.50\% &213.21 \\
& Qracle     &-0.80  &15.07\%  & 185.80 \\
& \textbf{$\Delta$}  
&\color{red}{1.27$\downarrow$}  
&\color{red}{6.43\%$\downarrow$}  
&\color{red}{27.41$\downarrow$} \\\hline
\multirow{4}{*}{\textbf{\ce{H2} Molecule}}
& Random     &-0.22  &\textcolor{blue}{2.01\%}  & \textcolor{blue}{215.91}\\
& Diffusion  &-0.36  &  58.17\%  &600 \\
& Qracle    &-0.58  &31.83\%&574.64 \\
& \textbf{$\Delta$}  
&\color{red}{0.22$\downarrow$}  
&\color{red}{26.34\%$\downarrow$} 
& \color{red}{25.36$\downarrow$}\\\hline
\multirow{4}{*}{\textbf{Random VQE}}
& Random     &0.02  & \textcolor{blue}{0.40\%} &\textcolor{blue}{237.82} \\
& Diffusion  &0.02  &0.80\%  &250.26 \\
& Qracle    &-0.08  & 0.40\% &247.56 \\
& \textbf{$\Delta$}  
& \color{red}{0.10$\downarrow$} 
&\color{red}{0.40\%$\downarrow$}   
&\color{red}{2.70$\downarrow$} \\\hline
\end{tabular}
\end{table}

\begin{figure*}[t]\centering
\begin{subfigure}[t]{0.48\textwidth}\centering
\includegraphics[width=\linewidth]{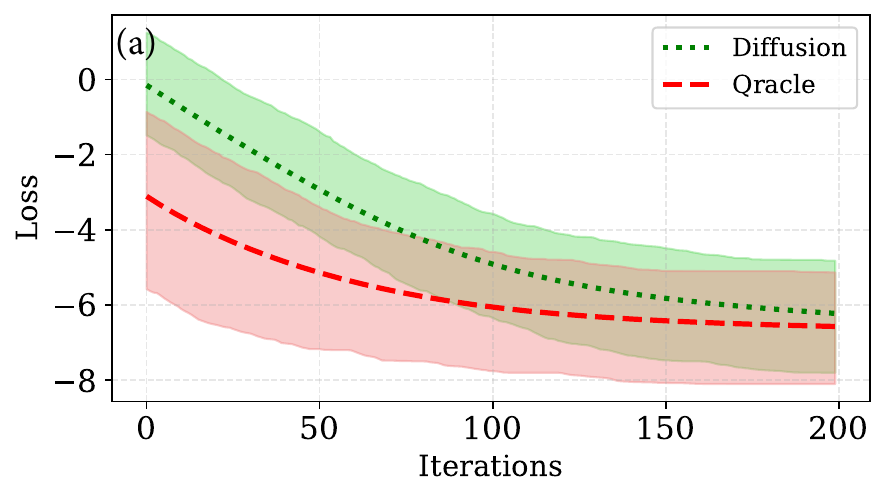}
\vspace{-0.2in}
\label{f:results_Heisenberg_loss}
\end{subfigure}
\hfill
\begin{subfigure}[t]{0.45\textwidth}\centering
\includegraphics[width=\linewidth]{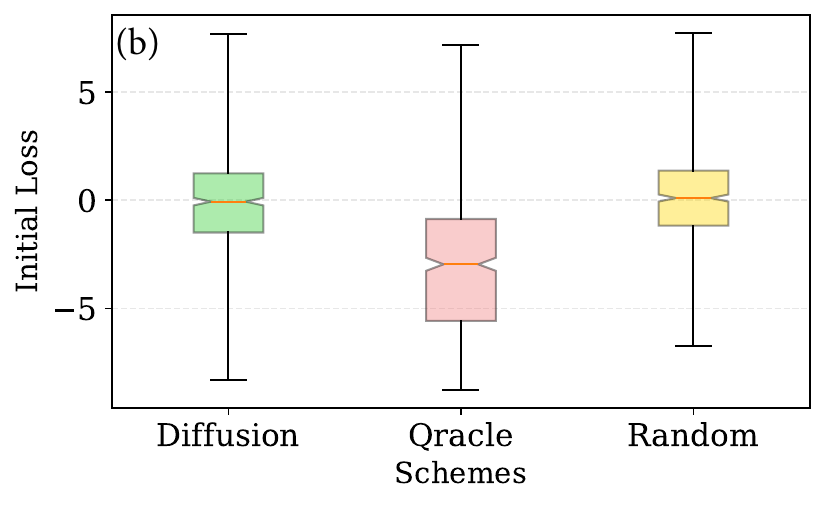}
\vspace{-0.2in}
\label{f:results_Heisenberg_distribution}
\end{subfigure}
\vspace{-0.10in}
\caption{Comparison of VQE parameter generation using~\design~and prior methods for the \textbf{\textit{Heisenberg XYZ model}}: (a) training loss trajectories and (b) initial loss distributions.}
\label{f:results_Heisenberg}
\vspace{-0.1in}
\end{figure*}
\begin{figure*}[t]\centering
\begin{subfigure}[t]{0.48\textwidth}\centering
\includegraphics[width=\linewidth]{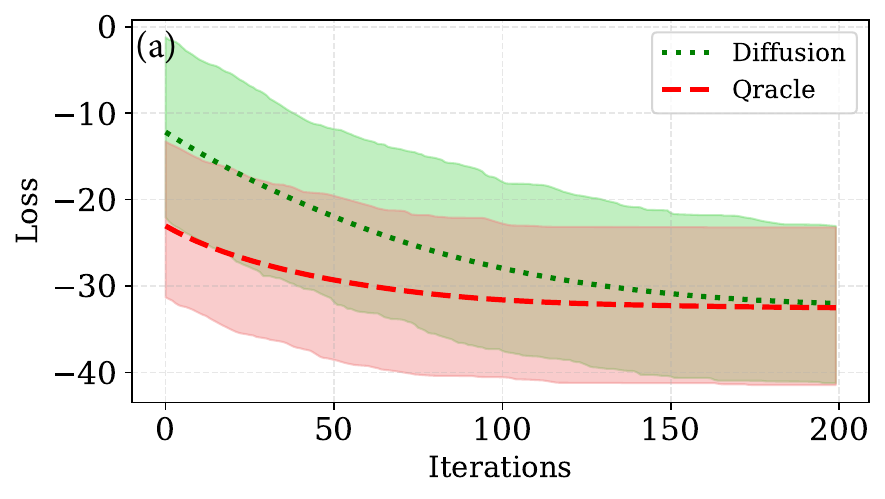}
\vspace{-0.2in}
\label{f:results_Ising_loss}
\end{subfigure}
\hfill
\begin{subfigure}[t]{0.45\textwidth}\centering
\includegraphics[width=\linewidth]{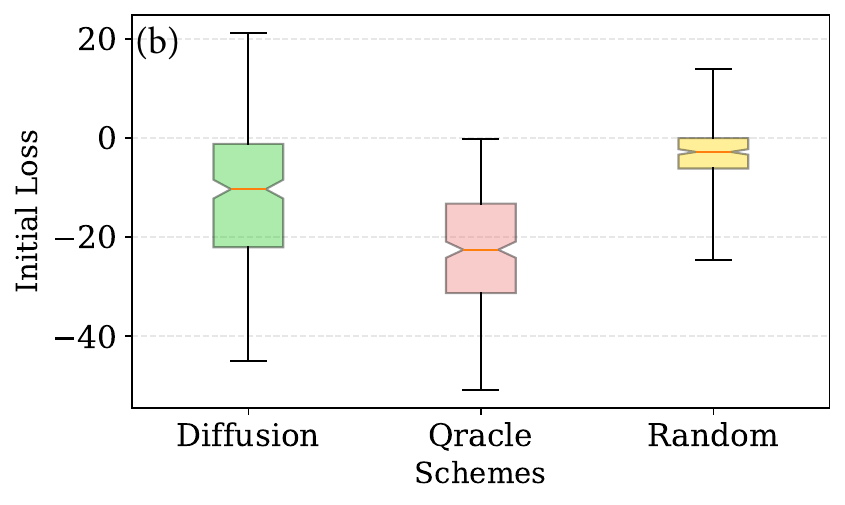}
\vspace{-0.2in}
\label{f:results_Ising_distribution}
\end{subfigure}
\vspace{-0.10in}
\caption{Comparison of VQE parameter generation using~\design~and prior methods for the \textbf{\textit{2D Ising model}}: (a) training loss trajectories and (b) initial loss distributions.}
\vspace{-0.1in}
\end{figure*}
\begin{figure*}[t]\centering
\begin{subfigure}[t]{0.48\textwidth}\centering
\includegraphics[width=\linewidth]{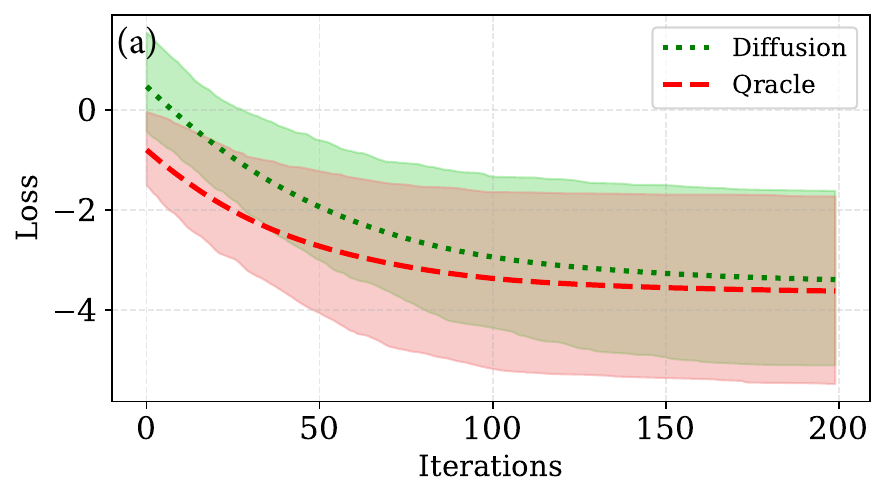}
\vspace{-0.2in}
\label{f:results_Fermi_loss}
\end{subfigure}
\hfill
\begin{subfigure}[t]{0.45\textwidth}\centering
\includegraphics[width=\linewidth]{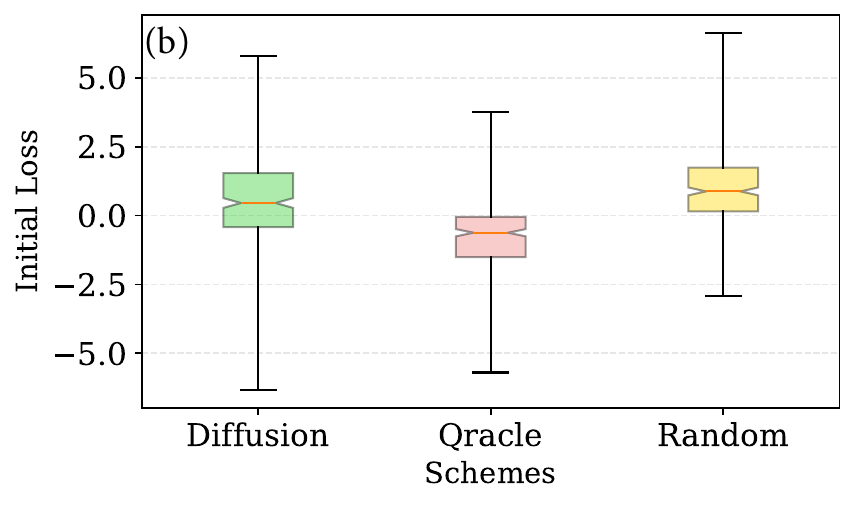}
\vspace{-0.2in}
\label{f:results_Fermi_distribution}
\end{subfigure}
\vspace{-0.10in}
\caption{Comparison of VQE parameter generation using~\design~and prior methods for the \textbf{\textit{Fermi--Hubbard model}}: (a) training loss trajectories and (b) initial loss distributions.}
\label{f:results_Fermi}
\vspace{-0.1in}
\end{figure*}

\textbf{Schemes}.  
To evaluate the performance of~\design~in comparison to established parameter initialization techniques for VQE circuits, we define the following two baselines:

\begin{itemize}[leftmargin=*, nosep, topsep=0pt, partopsep=0pt]
    \item \textit{Random}:  
    For each problem Hamiltonian and its VQE ansatz, parameters are randomly sampled from a standard normal distribution and assigned to the circuit. The resulting optimization trajectory and final performance are recorded.
    
    \item \textit{Diffusion}:  
    We implement the diffusion-based initializer proposed in~\cite{zhang2025diffusion}, where a latent diffusion model is trained on (\texttt{text\_prompt}, \texttt{opt\_param}) pairs from the training set of each application. For each problem Hamiltonian and its VQE ansatz, initialization parameters generated by the diffusion model are loaded into the circuit. The resulting optimization trajectory and final performance are recorded.
\end{itemize}

\textbf{Evaluation Metrics}.  
We use the following three metrics to evaluate performance across different schemes. These metrics capture complementary aspects of effectiveness—initialization quality, final solution accuracy, and optimization efficiency—and should be interpreted collectively, as no single metric alone fully represents overall performance.

\begin{itemize}[leftmargin=*, nosep, topsep=0pt, partopsep=0pt]
    \item \textit{Initial Loss}:  
    This metric reports the value of the loss function at the start of optimization for each scheme. It provides insight into how well the initializer positions the VQE circuit before training begins. While a lower initial loss may indicate a better starting point, it does not necessarily imply superior final performance.

    \item \textit{MRE or SMAPE}:  
    To evaluate final VQE performance, we compare predicted ground state energies against the true ground states of target Hamiltonians. In our preliminary study, we adopt Mean Relative Error (MRE) following~\cite{zhang2025diffusion} for consistency. For the final results in Section~\ref{sec:results}, we use Symmetric Mean Absolute Percentage Error (SMAPE)~\cite{SMAPE}, as MRE becomes unstable when the ground truth energy approaches zero—an edge case observed in some applications.

    \item \textit{Convergence Step}:  
    This metric quantifies the number of optimization steps required for the VQE circuit to converge. In this work, convergence is defined as the point at which the relative change in the loss function falls below a threshold of $10^{-5}$. The convergence step reflects both optimization efficiency and effectiveness of the initialization in accelerating convergence.
\end{itemize}

\begin{figure*}[t]\centering
\begin{subfigure}[t]{0.48\textwidth}\centering
\includegraphics[width=\linewidth]{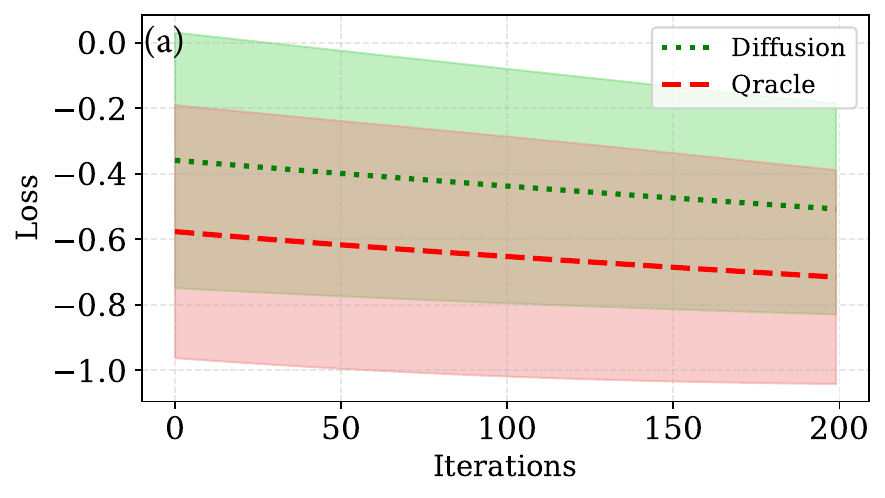}
\vspace{-0.2in}
\label{f:results_Fermi_loss}
\end{subfigure}
\hfill
\begin{subfigure}[t]{0.45\textwidth}\centering
\includegraphics[width=\linewidth]{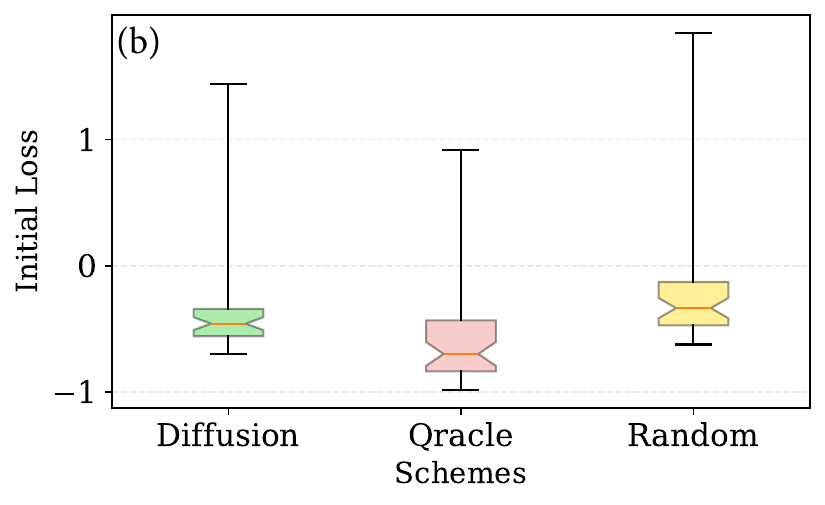}
\vspace{-0.2in}
\label{f:results_Fermi_distribution}
\end{subfigure}
\vspace{-0.1in}
\caption{Comparison of VQE parameter generation using~\design~and prior methods for the \textbf{\textit{Hydrogen molecule model}}: (a) training loss trajectories and (b) initial loss distributions.}
\label{f:results_H2}
\vspace{-0.1in}
\end{figure*}
\begin{figure*}[t]\centering
\begin{subfigure}[t]{0.48\textwidth}\centering
\includegraphics[width=\linewidth]{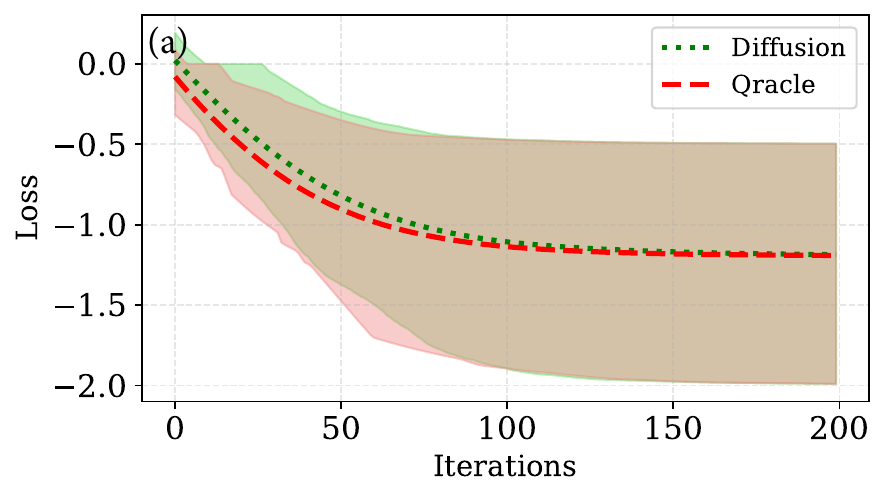}
\vspace{-0.2in}
\label{f:results_random_vqe_loss}
\end{subfigure}
\hfill
\begin{subfigure}[t]{0.45\textwidth}\centering
\includegraphics[width=\linewidth]{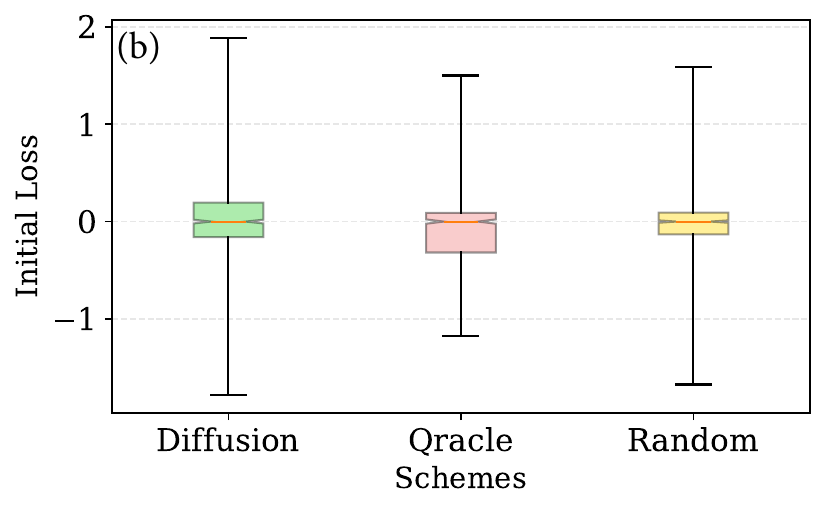} 
\vspace{-0.2in}
\label{f:results_random_vqe_distribution}
\end{subfigure}
\vspace{-0.1in}
\caption{Comparison of VQE parameter generation using~\design~and prior methods for the \textbf{\textit{random VQE model}}: (a) training loss trajectories and (b) initial loss distributions.}
\label{f:results_random_vqe}
\vspace{-0.1in}
\end{figure*}

\section{Evaluation and Results}
\label{sec:results}

\subsection{Overall Performance Comparison}

Table~\ref{tab:eval_metrics} presents a comprehensive performance comparison among~\design, randomly initialized VQE, and the diffusion-based model~\cite{zhang2025diffusion}. All values represent mean performance across all instances in the corresponding validation sets. 
We highlight the absolute performance improvements achieved by~\design~over the diffusion model~\cite{zhang2025diffusion} in red.
\design~consistently outperforms the SOTA diffusion based model~\cite{zhang2025diffusion} across all evaluation metrics.  More specifically, \design~achieves lower initial loss (with reductions of up to 10.86), faster convergence (requiring up to 64.42 fewer optimization steps), and improved final VQE performance (achieving up to a 26.43\% reduction in SMAPE). These improvements are particularly beneficial in the context of NISQ devices, where faster convergence directly translates to lower susceptibility to noise and decreased computational cost.

In comparison to random initialization, we find that in a few instances, it achieves performance comparable to—or even exceeding—that of \design on certain metrics, as highlighted in \textcolor{blue}{blue}. This underscores the stochastic nature of VQE optimization and suggests that, despite \design's overall robustness and generalizability, simpler heuristics may occasionally remain competitive in specific scenarios.

The superior performance of~\design~stems from two key factors. First, unlike\cite{zhang2025diffusion},   \design~employs a more expressive encoding by representing the Hamiltonian and ansatz topology as graphs, enabling task-specific representation learning. Second, its GNN architecture captures both local operator correlations via graph convolutions and long-range dependencies through attention mechanisms. These features enhance generalization and produce initialization parameters closer to optimal regions in the energy landscape.
This advantage is further evidenced by the per-application performance breakdown.~\design~shows the greatest improvements (e.g., a 26.34\% reduction in SMAPE) on complex Hamiltonians (e.g., \ce{H2} molecules), where rich physical structure and nontrivial entanglement patterns are present. In contrast, its benefits are less pronounced for random VQEs, which lack meaningful physical correlations. This disparity underscores~\design's ability to leverage domain-specific structure, making it particularly effective for realistic quantum applications that demand both scalability and precision.

\subsection{Application-Specific Results}
We also provide detailed learning curves of the loss function, which represents the expectation value of the Hamiltonian in the VQE setting, for both the diffusion-based model~\cite{zhang2025diffusion} and our method. In addition, we present the initial loss distributions across different initialization schemes. Specifically, all training loss trajectories shown in Figures~\ref{f:results_Heisenberg}–\ref{f:results_random_vqe} display pointwise mean values (dashed lines) across all test instances for each task, with shaded areas indicating the standard deviation.
As~\cite{zhang2025diffusion} focuses exclusively on Hamiltonians for many-body systems, we first compare its results with ours on such systems, and then extend the comparison to {\ce{H2}} molecules and the random VQE tasks introduced in this work.


\textbf{Quantum Many-Body Systems}.
As shown in Figs.~\ref{f:results_Heisenberg}–\ref{f:results_Fermi},  both \design~and the diffusion-based model achieve comparable final ground-state energies after 200 optimization steps, indicating similar VQE accuracy at convergence. However, \design~demonstrates significantly faster convergence and reduced optimization effort. For instance, in the 2D Ising model, \design~reaches an energy of $-32.01$ within just $123$ optimization steps, whereas the diffusion model requires 200 steps to attain the same energy level in average for all 300 circuits in the testing set. This highlights the training efficiency of \design.
Moreover, the initial loss distributions further underscore the advantage of \design. While the diffusion-based model sometimes exhibits similar initial loss values to random parameter initializations—as observed in the Heisenberg XYZ and Fermi-Hubbard models—it requires substantially more effort in data construction and training. In contrast, \design~consistently outperforms both random initialization and the diffusion-based model across all evaluated cases, demonstrating its robustness and superior initialization strategy for VQE tasks in many-body quantum systems.

\textbf{Hydrogen Molecule Model}.
The {\ce{H2}} molecular Hamiltonian is more complex than typical many-body systems due to its dense, non-local operator structure and more intricate optimization landscape. It arises from first-principles quantum chemistry and involves long-range Coulomb interactions. Moreover, fermionic-to-qubit mappings (e.g., Jordan–Wigner) introduce additional entanglement and circuit depth, and the need to capture electron correlation further increases the difficulty of VQE optimization.
As shown in the learning curves in Fig.~\ref{f:results_H2}(a), \design~achieves a significantly lower average initial loss ($-0.58$) compared to the diffusion-based model ($-0.36$) for all 45 circuits in the testing set, highlighting its enhanced ability to capture the nontrivial correlations present in molecular Hamiltonians. Although the initial loss distribution in Fig.~\ref{f:results_H2}(b) exhibits higher variance for \design, it remains concentrated in a lower-loss region overall. This indicates that \design~not only provides stronger average initialization performance but also yields more favorable starting points for subsequent VQE optimization.

\textbf{Random VQE Model}.
Fig.~\ref{f:results_random_vqe} presents the comparison across all schemes on randomly generated VQE tasks. Overall, \design~achieves only a marginal improvement in mean initial loss ($-0.08$) over the diffusion-based model ($0.02$), and their training trajectories remain largely similar. Additionally, the initial loss distributions for both \design~and the diffusion model show no significant advantage compared to random parameter initialization.
In contrast to the clear improvements observed for many-body systems and molecular Hamiltonians, these results reveal a key limitation of both machine learning-based approaches: they struggle to learn effective mappings from random Hamiltonians to optimized VQE ansatz parameters. This is likely due to the absence of meaningful physical structure or domain priors in the randomly generated Hamiltonians, which limits the models’ ability to generalize. These findings underscore that the success of learned initializations is strongly tied to the presence of underlying physical correlations in the Hamiltonian space.

\section{Conclusion}
This work presents \textit{Qracle}, a graph neural network (GNN)-based framework for VQE parameter initialization that addresses key limitations of existing methods, which often fail to capture the intricate correlations between the Hamiltonian and the ansatz circuit. By leveraging graph representations and GNN-based learning, \textit{Qracle} models these dependencies more effectively.
Extensive evaluations across diverse Hamiltonians show that \textit{Qracle} consistently outperforms state-of-the-art diffusion-based methods, achieving lower initialization error, faster convergence, and better final performance. Notably, it remains effective on complex systems where prior methods degrade.
These results highlight the promise of graph-based learning in variational quantum algorithms and provide a foundation for scalable, generalizable initialization strategies.

\section*{Acknowledgment}
This work was supported in part by NSF CCF-2105972, NSF OAC-2417589, and NSF CAREER AWARD CNS-2143120. 
We thank the IBM Quantum Researcher \& Educators Program for their support of Quantum Credits.
Any opinions, findings and conclusions or recommendations expressed in this material are those of the authors and do not necessarily reflect the views of grant agencies or their contractors.

\bibliographystyle{ieeetr}
\bibliography{quantum}

\end{document}